\documentclass[longauthor]{aastex701}
\shorttitle{Decline Relations}
\shortauthors{A. W. Shafter}
\submitjournal{Research Notes of the AAS}

\def\lessim{\mathrel{\hbox{\rlap{\hbox{\lower4pt\hbox{$\sim$}}}\hbox{$<$}}}}
\def\grtsim{\mathrel{\hbox{\rlap{\hbox{\lower4pt\hbox{$\sim$}}}\hbox{$>$}}}}
\usepackage{amsmath}

\begin{document}

\title{Modern Rate-of-Decline Relations for Novae}

\correspondingauthor{A. W. Shafter}

\author[0000-0002-1276-1486]{Allen W. Shafter}
\affiliation{Department of Astronomy and Mount Laguna Observatory, San Diego State University, San Diego, CA 92182, USA}
\email[show]{ashafter@sdsu.edu}

\begin{abstract}

A large sample of $t_2$ and $t_3$ times from the recent compilation of nova properties
given in Schaefer (2025) have been analyzed to determine relationships between these two parameters.
Fits were performed in both directions (from $\log t_2$ to $\log t_3$ and vice-versa)
to account for the asymmetry inherent in ordinary least-squares regression, which minimizes residuals only in the dependent variable.
The following best-fit relations were found: $\log t_3 = (0.877\pm0.019) \log t_2 + (0.444\pm0.027)$,
and $\log t_2 = (1.018\pm0.023) \log t_3 - (0.316\pm0.037)$, corresponding to
$t_3 = (2.78\pm0.17)~t_2^{(0.877\pm0.019)}$ and $t_2 = (0.483\pm0.041)~t_3^{(1.018\pm0.023)}$, respectively.
Within the uncertainties, the latter relation reduces to a simple proportionality: $t_2 \simeq 0.5~t_3$.

\end{abstract}

\keywords{\uat{Cataclysmic Variable Stars}{203} -- \uat{Novae}{1127} -- \uat{Recurrent Novae}{1366}}


\section{Introduction} 

The peak luminosities of novae and their rates of decline from maximum light provide important observational constraints on fundamental
system parameters such as the white dwarf mass and accretion rate \citep[e.g., see][]{Yaron2005,Shara2018,Shafter2026}. The custom
has been to parameterize the decline rate of novae as either the $t_2$ or $t_3$ time --
the time in days for a nova to decline by 2 and 3 mag from maximum light, respectively.
The $t_3$ parameter was first introduced by \cite{McLaughlin1945} in his seminal paper on the relation between a nova's peak luminosity
and its rate of decline (the MMRD relation). Roughly a decade later, \cite{Arp1956} adopted $t_2$
for analyzing novae in the Andromeda Galaxy (M31), as this was more practical for fainter, more
distant novae that often could not be followed 3~mag below peak brightness.

Transformations between $t_2$ and $t_3$ are often necessary when comparing the properties of novae where only one parameter
has been measured, and for comparing nova properties with models that require a specific input parameters.
Over the years, various empirical fits to available
nova data have been undertaken. Most notably, in his classic text \cite{Warner1995} used a sample of 52 Galactic novae
to show that $t_3 \simeq 2.75~t_2^{0.88}$. Another empirical relation was offered
by \cite{Capaccioli1990} who found $t_3=1.68~(\pm 0.08) + 1.9~(\pm 1.5)$ for $t_3<80$~d and $t_3=1.68~(\pm0.04) + 2.3~(\pm1.6)$ for $t_3>80$~d.
The number of nova light-curves available for study has increased significantly in the
more than three decades that have elapsed since the publication of these relations.
In this Research Note, I present modern re-determinations of the nova decline-rate relations,
which incorporate the most recent nova light-curve data.

\begin{figure}
\plotone{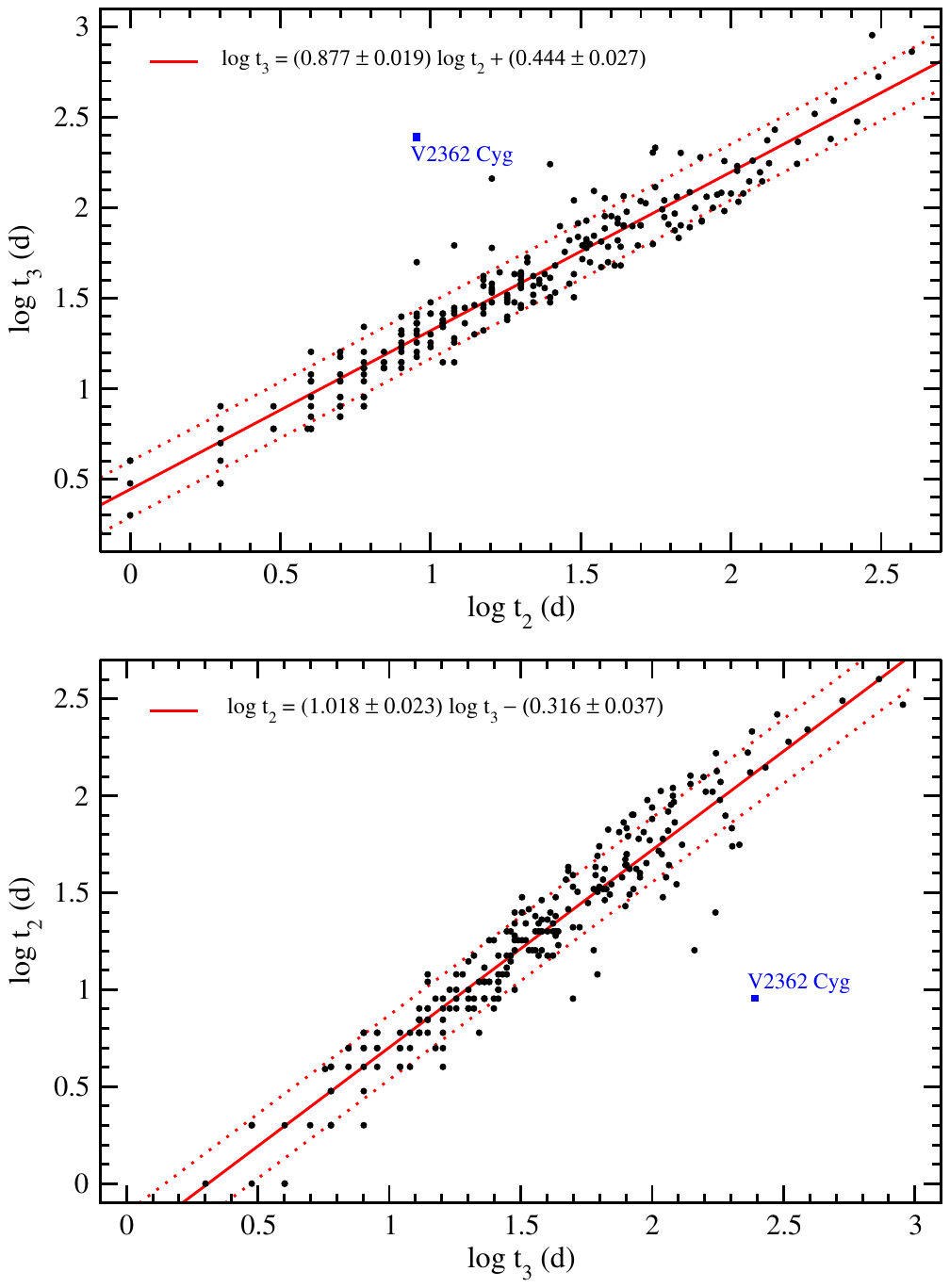}
\caption{Top panel: The linear least-squares fit of $\log t_3$ as a function of $\log t_2$ for 244 novae included in the
\citet{Schaefer2025} nova sample. The best fitting linear relation is shown by the solid red line.
Bottom Panel: The least-squares fit of $\log t_2$ as a function of $\log t_3$ for the same sample of novae. Again,
the best fitting linear relation is shown by the red line.
In both panels, the dotted red lines show the $\pm1\sigma$ error bands to the fit, while
the position of the unusual system, V2362~Cyg, is shown in blue.}
\end{figure}

\section{Updated Rate-of-Decline Relations for Novae}

In a series of papers, B.E. Schaefer has assembled an unprecedented wealth of observational data on Galactic
novae \citep{Strope2010,Schaefer2022,Schaefer2025}. In particular, Table~1 of \citet{Schaefer2025} contains a wide range of
observed and derived properties for a total of 402 Galactic novae.
Among these properties include measurements for {\it both\/} the $t_2$ and $t_3$ times for 245 novae\footnote{One system,
MT~Cen, which erroneously shows $t_2 > t_3$, has been omitted from the present analysis.}.

A key aspect of the analysis is performing independent least-squares regression fits in both directions:
one treating $\log t_2$ as the independent variable to predict $\log t_3$, as was done by \cite{Warner1995} and \cite{Capaccioli1990},
and the other treating $\log t_3$ as independent variable to predict $\log t_2$.
This approach is necessary because ordinary least-squares regression is inherently
asymmetric. It minimizes the vertical residuals in the dependent variable only, leading to different fits when the axes are swapped.
Consequently, the power-law exponents are not exact reciprocals, due to intrinsic scatter in both $t_2$ and $t_3$
arising from diverse light-curve morphologies (e.g., dust dips, plateaus, or oscillations) and observational uncertainties.

The results of the study are shown in Figure~1. The top panel shows that $\log t_3 = (0.877\pm0.019) \log t_2 + (0.444\pm0.027)$, or
$t_3 = (2.78\pm0.17)~t_2^{(0.877\pm0.019)}$. The data fit a power-law relation remarkably well, with the exception of the unusual
nova V2362~Cyg, which exhibited a rapid initial drop in brightness followed by a prolonged rebrightening phase \citep{Kimeswenger2008}.
It is remarkable that the best-fitting relation is virtually identical
to that found previously by \citet{Warner1995} in his classic study of just 52 novae.
The bottom panel shows that $\log t_2 = (1.018\pm0.023) \log t_3 - (0.316\pm0.037)$, demonstrating that
the best-fit solution is not simply the inverse of the previous fit. Instead, I find
$t_2 = (0.483\pm0.041)~t_3^{(1.018\pm0.023)}$, which, within the uncertainty in the fit, reduces to: $t_2 \simeq 0.5~t_3$.
Simply inverting Warner's relation $(t_2 \simeq [t_3 / 2.75]^{1/0.88})$ would overestimate the exponent (1.14) compared to
that for a direct fit (1.02), potentially introducing biases of $\sim$15\% in the predicted $t_2$ values
for the fastest and slowest novae.

\section{Summary}

Least-squares fits of the $t_2$ and $t_3$ times for 244 novae listed in Table~1 of \citet{Schaefer2025} have been performed.
When estimating $t_3$ from $t_2$, the best-fit solution yields:

\begin{equation}
t_3 = (2.78\pm0.17)~t_2^{(0.877\pm0.019)},
\end{equation}

\noindent
which is equivalent (within the uncertainties) to the well-known relation found previously by \cite{Warner1995}.
On the other hand, when transforming from $t_3$ to $t_2$, I find:

\begin{equation}
t_2 = (0.483\pm0.041)~t_3^{(1.018\pm0.023)},
\end{equation}

\noindent
which, within the uncertainties, can be approximated by a simple proportionality: $t_2 \simeq 0.5~t_3$.

In cases where uncertainties in the inputs ($\sigma_{t_{2,i}}, \sigma_{t_{3,i}}$) are known,
standard propagation of errors yields:

\begin{subequations}
\begin{equation}
\sigma_{t_3} = t_3 \times \sqrt{0.00387 + (0.877 \sigma_{t_{2,i}}/t_2)^2 + (0.0438 \log t_2)^2},
\end{equation}
\begin{equation}
\sigma_{t_2} = t_2 \times \sqrt{0.00726 + (1.018 \sigma_{t_{3,i}}/t_3)^2 + (0.0530 \log t_3)^2}.
\end{equation}
\end{subequations}

Finally, I note that even among S-type (smooth) novae, the forward and reverse slopes remain significantly different (though less pronounced),
indicating that the directional dependence reflects intrinsic scatter and slight non-linearity, not only mixing of morphological classes.

\bibliography{novarefs}{}
\bibliographystyle{aasjournalv7}

\end{document}